\documentclass[twocolumn,showpacs,preprintnumbers,amsmath,amssymb,%
nofootinbib,floatfix]{revtex4}

\usepackage{graphicx}
\usepackage{dcolumn}
\usepackage{bm}
\usepackage{subfigure}
\usepackage{ifthen}
\begin{document}

\title{Description of nuclear octupole and quadrupole deformation
close to the axial symmetry:
Critical-point behavior of 
$^{224}$Ra  and $^{224}$Th}

\author{P.G. Bizzeti}
\email{bizzeti@fi.infn.it}
\author{A.M. Bizzeti--Sona}
\affiliation{Dipartimento di Fisica, Universit\'a di Firenze\\
I.N.F.N., Sezione di Firenze\\
Via G. Sansone 1, 50019 Sesto Fiorentino (Firenze), Italy}
\date{\today}

\begin{abstract}
The model, introduced in a previous paper, for the description
 of the octupole and quadrupole degrees of
freedom in conditions close to the axial symmetry, is 
applied to situations of shape phase transitions where the 
quadrupole amplitude can reach zero.
The transitional nuclei $^{224,226}$Ra and $^{224}$Th are discussed
in the frame of this model. Their level schemes can be reasonably 
accounted for assuming a square-well potential in two dimensions. 
Electromagnetic transition amplitudes are also evaluated and 
compared with existing experimental data.  
\end{abstract}


\maketitle

\section{Introduction}
\label{S:1}
The phase transition  between spherical and axially deformed quadrupole
 shape of nuclei has been the object of several theoretical and
experimental works in recent years. In particular, the properties of
nuclei close to the critical point, predicted by Iachello's model of
X(5) symmetry~\cite{iachello01}, have been actually observed in several 
cases~\cite{casten01,kruecken02,bizzeti02,clark03,hutter03,tonev04}, 
while some other 
nuclides showing the ratio $E(4+)/E(2^+)\approx 2.91$
expected for the X(5) symmetry are presently under 
investigation. Moreover,
in the Ra -- Th region, it has been observed that the isotopes
$^{224}$Ra and $^{224}$Th have a positive--parity ground--state band
with a sequence of level energies very close to the X(5) 
predictions~\cite{bizzetierice,bizzetirila}.
Here, however, the presence of a very low lying negative parity band,
soon merging with the positive--parity one for $J>5$, proves that
the octupole mode of deformation plays an important role and should
not be ignored in discussing the behavior of the phase transition.

In a previous paper~\cite{bizzeti04} (henceforth referred to as I) 
a simple model has been introduced to describe the phase transitions
in nuclear shape involving the octupole mode\footnote{We have now the
occasion to correct a few misprints which escaped proofs revision in 
paper I:\\
Eq. (23c) should read\ $q_3= \left( L_3-p_\varphi-2p_\chi
-3p_\vartheta \right) \ / \left( 4 u_0^2 \right)\ .$\\
In the Table VII, the 4th element of the 5th line should be
${\sin \theta_2 \sin \theta_3}\ / \ {\mathcal{J}_1}\ $.
We apologize for these errors.}. To this purpose, a new
parametrization of the collective coordinates describing the nuclear
quadrupole and octupole deformation has been introduced and discussed.
The nuclear shape is represented in the intrinsic frame defined by the
principal axes of the overall tensor of inertia, in situations close
to (but not necessarily coincident with) the axial--symmetry limit.
In the same paper, a specific model is developed to describe the
critical point of the phase transition in the octupole mode, between
harmonic oscillations and permanent asymmetric deformation, in nuclei
which already possess a stable quadrupole deformation. The Thorium
isotopic chain was investigated and the
experimental data concerning $^{226,228}$Th were compared with the
model predictions~\cite{bizzeti04}. The former appears to be close to
the critical point, while the latter can be interpreted as an example
of harmonic oscillations in the axial octupole mode.

In the present paper, we extend the investigation 
to the cases where
the quadrupole deformation is not steady but performs oscillations
under the effect of a proper potential, and in
particular for situations close to the quadrupole critical point 
described by the X(5) symmetry, in the Radium and Thorium 
isotopic chain. 

As we shall see, the properties of the 
already mentioned nuclei $^{224}$Th and $^{224}$Ra result to be 
reasonably
described by our model with a ``critical'' (flat) potential well,
extending both in the $\beta_2$ and $\beta_3$ directions.
Moreover, we observe that, as far as the level scheme is concerned,
 also the next isotope $^{226}$Ra 
can be accounted for with a proper critical--point potential, 
in spite of the fact that the positive--parity part of
the ground--state band does not follow the X(5) predictions. 
As in the case of Thorium, heavier isotopes of Radium have a 
permanent quadrupole deformation and octupole
excitations of vibrational character, while the 
lighter ones are either  non collective or vibrational in the 
quadrupole mode. 

Some results of this work, at different phases of advancement, 
have been reported at several Conferences or 
Schools~\cite{bizzetirila,bizzeticamerino,bizzetipredeal,bizzetivico}.

For convenience of the reader, we report in the next 
Section~\ref{S:2} some evidence of the phase transitions
in the Radium and Thorium isotopic chain, while the definition of 
variables introduced in I, and a few results relevant to the present
work,  are briefly summarized in the Section~\ref{S:3a}. 
In the following subsections, the model introduced in I is specialized 
to a form suitable for a critical potential in two dimensions. 
Finally, in the Section~\ref{S:4}
the model results are reported and compared with the existing 
experimental evidence
for $^{224,226}$Ra  and $^{224}$Th.

Previous models of quadrupole--octupole deformation are quoted in I. 
Since then, new relevant papers have appeared. 
A new ``Analytic Quadrupole--Octupole axially symmetric model''
(AQOA) has been proposed by D.Bonatsos {\it et al.} \cite{bonatsos05}
to discuss the evolution
of the quadrupole and octupole collectivity in Ra and Th isotopes.
A parameter free model starting from a similar approach has
been developed by Lenis and Bonatsos~\cite{lenis06} and compared with
the experimental results for $^{226}$Ra and $^{226}$Th. 
Moreover, a variant of the AQOA model, introducing
a renormalization of the nuclear moment of inertia~\cite{minkov06},
has been used to describe the lowest quadrupole and octupole
bands of the $N=90$ isotones $^{150}$Nd,  $^{152}$Sm, $^{154}$Gd and
$^{156}$Dy. A discussion of the octupole bands of $^{150}$Nd and
$^{152}$Sm is also contained in the 
Ref.s~\cite{bizzetirila,bizzetipredeal}. Finally, 
an extension of the Extended Coherent State model~\cite{raduta03} has 
been developed by 
A.A. Raduta and coworkers~\cite{raduta06b,raduta06c,radutapredeal}, 
to include in the model space also the lowest $K^\pi=1^+$ and $1^-$ bands,
and the model predictions have been compared with the experimental
data for several nuclei of the  regions of the  actinides and of 
the rare-earths.

\section{\label{S:2}Phase transitions in the R\lowercase{a} -- 
T\lowercase{h} region}

We summarize here the existing evidence for the evolution of
nuclear shapes for Radium and Thorium isotopes in the transitional 
region $N=130 - 140$.

\begin{figure}[!th]
\includegraphics[height=8.cm,clip,angle=90,bb=79 20 517 750]
{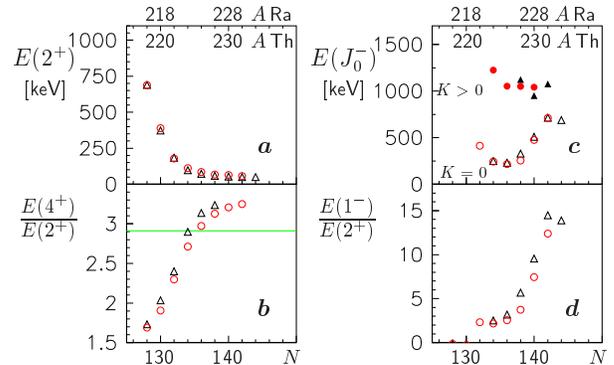}
\caption{\label{F:1} (Color online) Indicators of the quadrupole 
collectivity (left)
and of the octupole collectivity (right), as a function of the neutron
number $N$ in the isotopic chain of Ra (circles) and Th (triangles):
$(a)$ - Excitation energy of the first $2^+$ level; $(b)$ - Energy ratio
$E(4^+)/E(2^+)$; $(c)$ - Excitation energy of the first level of the
$K^\pi=0^-$ band, $J_0^\pi=1^-$ (open symbols) and of the lowest known
level of other negative-parity bands, $J_0^\pi=2^-$ or $1^-_2$ (full
symbols); $(d)$ - Energy ratio $E(1^-)/E(2^+)$.
The horizontal line in the part $(b)$ shows the value (2.91) 
expected for the X(5) symmetry (from ref.~\cite{bizzeti04}). }
\end{figure}

The Fig.~\ref{F:1}, taken from I, shows the behavior of some
indicators of quadrupole and octupole collectivity, as a 
function of the neutron number $N$, in the isotopic chain of
Ra and Th. It has been noted in I that $^{226}$Th appears to
be close to the critical point in the octupole deformation,
while it possesses a stable quadrupole deformation $\beta_2$.
At larger values of $N$, Th isotopes maintain a stable quadrupole
deformation, while the octupole mode evolves towards
the vibrational behavior, as indicated by the large excitation 
energies of all negative--parity 
levels. At $N=130$ or less, the quadrupole mode has
a vibrational (or non collective) character. 
It turns out, therefore, that the octupole phase transition
 proceeds in the direction opposite to the one of quadrupole.
We also observe that the phase transition only involves the 
axial octupole mode.
In fact, the energy of the $J^\pi=1^-$ band
head of the $K^\pi=0^-$ octupole band shows a sharp decrease, both in
its absolute value and in the ratio to the $E(2^+)$, when the neutron
number decreases below $N=142$. Other octupole bands (with $K>0$) do
not show a similar trend (fig.~\ref{F:1}$(c)$), and one can conclude
that non-axial octupole excitations maintain a vibrational character.
 A similar trend is apparent also for Ra isotopes.

In order to describe Th and Ra isotopes with $A < 226$, we must
allow also $\beta_2$ to vary and perform (non necessarily harmonic)
oscillations. If we consider the value
$E(4^+)/E(2^+)=2.91$ as a signature of the critical point with
respect to the quadrupole deformation, this would correspond 
approximately to $^{224}$Ra and $^{224}$Th. 

\noindent

\section{The model for quadrupole -- octupole vibrations}
\subsection{\label{S:3a}Summary of the variable definitions}

The dynamical variables $a^{(\lambda )}_{\mu}$ ($\lambda=2,3;\ \mu
=-\lambda ... \lambda$), describing the quadrupole and octupole
deformation in the intrinsic reference frame, are parametrized as
\begin{eqnarray}
a^{(2)}_0 &=& \beta_2 \cos \gamma_2 \approx \beta_2 \nonumber\\*
a^{(2)}_1 &=& -  \frac{\sqrt{2}\ \beta_3}{\sqrt{\beta_2^2+2 
\beta_3^2}}\ v\ (\sin \varphi + i \cos \varphi )\cr 
a^{(2)}_2 &=& \sqrt{1/2}\ \beta_2\ \sin \gamma_2 - i 
\frac{\sqrt{5}\ \beta_3}{\sqrt{\beta_2^2+2 \beta_3^2}}\ u \sin \chi  \cr
a^{(3)}_0 &=& \beta_3 \ \cos \gamma_3 \approx \beta_3\cr
a^{(3)}_1 &=&\frac{\sqrt{5}\ \beta_2}{\sqrt{\beta_2^2+2 \beta_3^2}}\ v
\  (\sin \varphi + i \sin \varphi ) \cr
a^{(3)}_2 &=& \sqrt{1/2}\ \beta_3\ \sin \gamma _3 +i
\frac{\beta_2}{\sqrt{\beta_2^2+2 \beta_3^2}}\  u \sin \chi \cr
a^{(3)}_3 &=& w \sin \vartheta \left[ \cos \gamma_3 + 
(\sqrt{15}/2)\ \sin \gamma_3 \right]\cr
  &+& i\ w \cos \vartheta \left[ \cos \gamma_3 - 
(\sqrt{15}/2)\ \sin \gamma_3 \right] \cr
&\approx& w\ (\sin \vartheta +i \cos \vartheta )
\label{E:2.2}
\end{eqnarray}
With this choice, valid in situations close to the axial symmetry,
the tensor of inertia turns out to be diagonal {\em up to the first order}
in the {\em small} quantities describing the non-axial deformations. 

In the Eq.s~\ref{E:2.2} the variables $\gamma_2$ and $\gamma_3$ are still
employed, in order to keep some transparency with respect to the standard 
expressions used to describe the quadrupole~\cite{bohr52}
 or the octupole deformation alone~\cite{wexler99}. 
However, it is more convenient to substitute
them with expressions involving the variables $u, \chi$ and 
a new variable $u_0$: 
neglecting second-order and higher-order terms,
\begin{eqnarray}
\gamma_2 &=& \frac{\sqrt{10}\beta_3}{\beta_2
\sqrt{\beta_2^2+5\beta_3^2}} u \cos \chi 
+\frac{f(\beta_2,\beta_3)}{\sqrt{\beta_2^2+5\beta_3^2}} u_0 \\*
\gamma_3&=&-\frac{\sqrt{2}\beta_2}{\beta_3
\sqrt{\beta_2^2+5\beta_3^2}} u \cos \chi 
+\frac{\sqrt{5} f(\beta_2,\beta_3)}{\sqrt{\beta_2^2+5\beta_3^2}} u_0 \nonumber
\label{E:2.3}
\end{eqnarray}
It is possible to show that a definite value of the angular-momentum 
component $K$ along the
intrinsic axis 3 and a definite parity
can be associated to the degrees of freedom corresponding to
the variables $v, \chi$ (or $u, \varphi$ or $w, \vartheta$ or $u_0$):
 $K^\pi=1^-$ (or $2^-$ or $3^-$ or $2^+$, respectively). 
This result is independent of the form of the function 
$f(\beta_2, \beta_3)$ of Eq.~\ref{E:2.3} (which, actually, was left 
undetermined in I).

\subsection{The kinetic energy operator}
The classical expression of the kinetic energy has the form
\begin{equation}
T=\frac{1}{2}\sum \mathcal{G}_{\mu \nu}\dot{\xi}_\mu\dot{\xi}_nu
\label{E:2.6}
\end{equation}
where $\dot{\xi}\equiv (\dot{\beta}_2,\dot{\beta_3},\dot{u_0},
 \dot{v},\dot{\chi},\dot{u},\dot{\varphi},\dot{w},\dot{\vartheta},
q_1,q_2,q_3)$ and $q_1$, $q_2$, $q_3$ 
are the components of the angular velocity
along the three axes of the intrinsic reference frame.  As in I, we
adopt here the convention of including the inertial coefficient
$B_\lambda$ in our amplitudes $a^{(\lambda)}_\mu$, that therefore 
would correspond to $\sqrt{B_\lambda}a^{(\lambda)}_\mu$ in the usual
notations of Bohr. 
\begin{table}[bt]
\caption{\label{T:1}  The  matrix of inertia $\mathcal{ G}$ 
after the introduction of the variables
$u_0$, $v$, $u$,
$w$,  $\varphi$,  $\chi$, and $\vartheta$ (see text).
Here, $\mathcal{ J}_1 = \mathcal{ J}_2 = 3(\beta_2^2 + 2\beta_3^2)$, and
$\mathcal{ J}_3 = 4 f^2(\beta_2,\beta_3)
\ u_0^2 + 
2v^2 + 8u^2 + 18w^2$.
Only the leading terms are shown. Neglected terms are small of the
first order (or smaller) in the sub-matrix involving only 
 $\dot{\beta_2}$, $\dot{\beta_3}$, $\dot{u_0}$, 
$\dot{v}$, $\dot{u}$, $\dot{w}$, 
 $q_1$ and $q_2$; of the third order (or smaller) in the sub-matrix 
involving only
$\dot{\varphi} $, $\dot{\chi}$,  $\dot{\vartheta}$ and $q_3$;
of the second order (or smaller) in the rest of the matrix.}
\begin{ruledtabular}
\begin{tabular}{l|ccccccccc|ccc}
& $\dot{\beta_2}$ & $\dot{\beta_3}$ & $\dot{u_0}$ & 
$\dot{v}$ & $\dot{u}$ & $\dot{w}$ 
& $\dot{\varphi} $ & $\dot{\chi}$ &  $\dot{\vartheta}$ & $q_1$
& $q_2$  & $q_3$\\
\hline
$\dot{\beta_2}$ & 1 & 0 & 0 & 0 & 0 & 0 & 0 & 0 & 0 & 0 & 0 & 0 \\
$\dot{\beta}_3$ & 0 & 1 & 0 & 0 & 0 & 0 & 0 & 0 & 0 & 0 & 0 & 0 \\
$\dot{u_0}$ & 0 & 0 & $f^2(\beta_2,\beta_3)$
%
 & 0 & 0 & 0 & 0 & 0 & 0 & 0 & 0 & 0 \\
$\dot{v}$    & 0 & 0 & 0 & 2 & 0 & 0 & 0 & 0 & 0 & 0 & 0 & 0 \\
$\dot{u}$   & 0 & 0 & 0 & 0 & 2 & 0 & 0 & 0 & 0 & 0 & 0 & 0 \\
$\dot{w}$       & 0 & 0 & 0 & 0 & 0 & 2 & 0 & 0 & 0 & 0 & 0 & 0 \\
$\dot{\varphi}$ & 0 & 0 & 0 & 0 & 0 & 0 & $2v ^2$   & 0 & 0 
                                           & 0 & 0 &  $2v ^2$ \\
$\dot{\chi}$    & 0 & 0 & 0 & 0 & 0 & 0 & 0 & $2u^2$ & 0 
                                    & 0 & 0 &  $4u^2$ \\
$\dot{\vartheta}$ & 0 & 0 & 0 & 0 & 0 & 0 & 0 & 0 & $2 w^2$
                                           & 0 & 0 & $6 w^2$ \\
\hline
$q_1$ & 0 & 0 & 0 & 0 & 0 & 0 & 0 & 0 & 0 & $\mathcal{ J}_1$ & 0 & 0 \\
$q_2$ & 0 & 0 & 0 & 0 & 0 & 0 & 0 & 0 & 0 & 0 & $\mathcal{ J}_2$ & 0 \\
$q_3$ & 0 & 0 & 0 & 0 & 0 & 0 & $2v^2$ & $4u^2$ & $6w^2$
                                          & 0 & 0 & $\mathcal{ J}_3$ \\
\end{tabular}
\end{ruledtabular}
\end{table}
The matrix
elements of $\mathcal{G}$, approximated to the most relevant order, 
are shown in the Table~\ref{T:1}. 
The determinant of this matrix turns out to be
\begin{eqnarray}
G &\propto & (\beta_2^2 + 2 \beta_3^2)^2\ f^4(\beta_2, \beta_3) 
\ u_0^2 \ v^2 u^2 w^2\nonumber \\[2pt]
&\equiv & G_0(\beta_2,\beta_3)\ u_0^2 \ v^2 u^2 w^2 .
\label{E:2.7}
\end{eqnarray}
The Pauli recipe for the quantization of the classical kinetic energy
gives the Schr\"odinger equation
\begin{equation}
\sum_{\mu \nu} \frac{1}{g} 
\frac{\partial}{\partial \xi_\mu} \left[ g 
\!\left( \mathcal{G}^{-1} \right) _{\mu \nu} 
\frac{\partial \Psi}{\partial \xi_\nu }
\right]+ \frac{2}{\hbar^2} \left[ E\!-\!V(\xi ) \right] \Psi\! =\!0
\label{E:a1.2}
\end{equation}
where $g^2=G={\rm Det}\ \mathcal{G}$ and $\xi$ stays for the ensemble of 
the variables $\xi_\kappa$.

To  our present purpose, this general treatment must be 
specialized, ({\it e.g.} with a proper choice of the arbitrary 
function $f(\beta_2,\beta_3)$ in the Eq.s~\ref{E:2.3}), 
keeping in mind a necessary condition:  the
Schr\"odinger equation for the quadrupole amplitude, when the octupole
amplitude is constrained to small values by a proper restoring
potential, must converge to that of Bohr and therefore, at 
the critical--point, to that of the X(5) 
model\footnote{This choice is
different from the one adopted in I to describe the critical point in
the octupole degree of freedom with a constant quadrupole deformation:
in such a case, in fact, the proper limit for small octupole 
amplitudes does not correspond to the X(5) but to the 
Frankfurt model ~\cite{eisen1}, valid for small--amplitude octupole 
vibrations of a well deformed nucleus.}.
As we shall see in the Section~\ref{S:4.1}, this result is obtained
with the choice
\begin{equation}
f(\beta_2,
\beta_3)=\sqrt{\frac{(\beta_2^2+\beta_3^2)(\beta_2^2+2\beta_3^2)}
{\beta_2^2+5\beta_3^2}}\ ,
\label{E:2.4}
\end{equation}
from which one obtains
\begin{equation}
G_0(\beta_2,\beta_3)=
\frac{(\beta_2^2+\beta_3^2)^2(\beta_2^2+2\beta_3^2)^4}
{(\beta_2^2+5\beta_3^2)^2}\ ,
\label{E:2.4b}
\end{equation}
\subsection{The critical potential in two dimensions}
\begin{center}
\begin{figure}[t]
\includegraphics[width=8.4cm,clip,bb=14 450 537 728]
{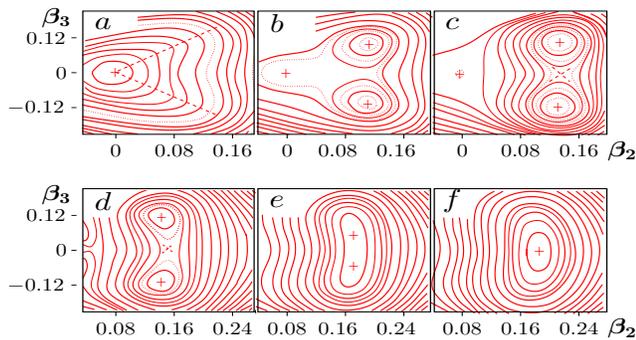}
\caption{(color on line) Potential--energy surfaces in the 
$\beta_2$ -- $\beta_3$
plane for several Th isotopes, as given by Nazarewicz 
{\it et al.}~\cite{nazarew85}.}
\label{F:2}
\end{figure}
\end{center}

\begin{center}
\begin{figure}[h]
\includegraphics[height=5.9cm,clip,angle=90,bb=205 485 425 815]
{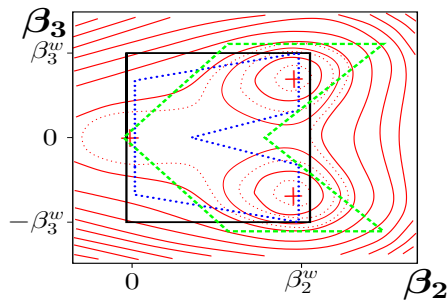}
\caption{(color on line) Possible shapes for a potential well 
simulating the critical--point potential. The potential--energy 
surface of fig.~\ref{F:2}$(b)$ is also shown for comparison.}
\label{F:3}
\end{figure}
\end{center}

Possible landscapes of axial quadrupole--octupole deformation in the
Thorium region are exemplified in Fig.~\ref{F:2}, where the 
potential energy is depicted as a function of the deformation 
parameters $\beta_2$ and $\beta_3$.  Reported values have been 
obtained by Nazarewicz {\it et al.}~\cite{nazarew85} with a 
Wood--Saxon--Bogolyubov cranking calculation. We notice
that fig~\ref{F:2}$(e)$ shows a potential minimum which is localized 
around a fixed value in the  $\beta_2$ direction, while a flat minimum
extends over a sizable interval in the direction $\beta_3$. This is
just the ``critical'' potential for the shape transition between
octupole oscillation and permanent octupole deformation (combined with
a fixed quadrupole deformation), corresponding to the 
fig.s~\ref{F:2}$(d)$ and \ref{F:2}$(f)$, respectively.
The fig.s~\ref{F:2}$(a), \ref{F:2}(b)$ and \ref{F:2}$(c)$, 
instead, show a different 
kind of shape transition, proceeding directly from a fixed, 
reflection--asymmetric deformation (fig.\ref{F:2}$(c)$) to
quadrupole--octupole vibrations around a spherical shape
(fig.\ref{F:2}$(a)$). 
The potential corresponding to the critical point is not shown.
It should be somewhere between fig.~\ref{F:2}$(a)$ and
fig.~\ref{F:2}$(b)$.
 One can try to approximate the critical 
potential, as usual, with a square well,
but now the flat bottom of the well should extend 
over a finite distance in $\beta_2$ and $\beta_3$,
and be symmetric in $\beta_3 $ around $\beta_3=0$.
The shape of the borders is obviously relevant to the result.
One could imagine shapes like those shown in fig.~\ref{F:3} with 
dashed or dotted lines, but their description would involve at least 
two or three free parameters, and the comparison with experimental
data could be not very significant. We have found, however, that
good results are obtained also
 with a simple rectangular shape
(solid line in fig.~\ref{F:3}), implying only one free parameter,
$b=\beta_3^w/\beta_2^w$ (apart from a common factor of scale).
\section{\label{S:4}Results and comparison with experimental data}
\subsection{\label{S:4.1} The Energy eigenvalues}
\noindent

\begin{figure*}[!t]
\begin{center}
\includegraphics[height=\linewidth,angle=90,clip,bb=260 100 485 835]
{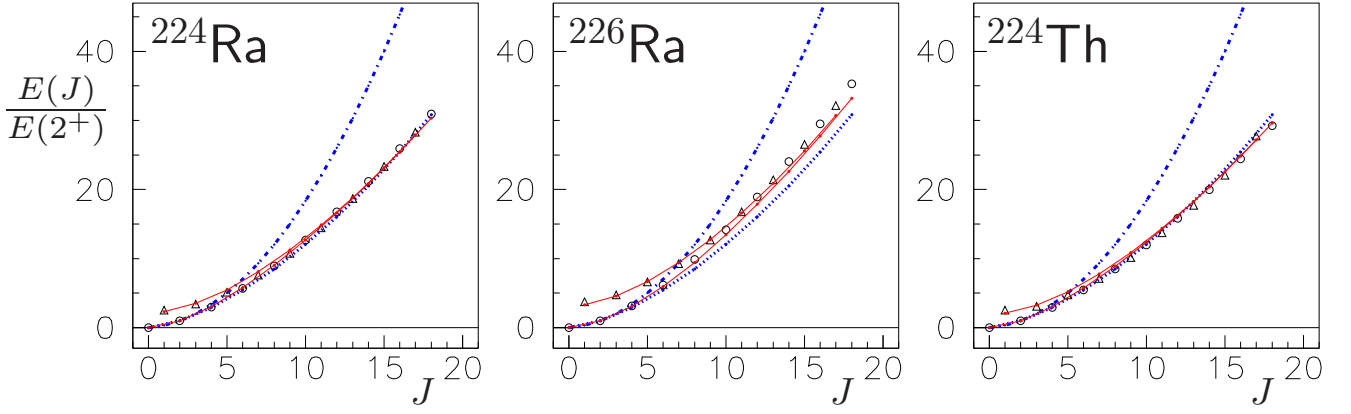}
\end{center}
\caption{(Color online) Experimental excitation energies of the 
positive
parity levels (circles) and of the negative parity ones (triangles),
in units of $E(2^+)$,
for $^{224,226}$Ra and $^{224}$Th, compared with the results of the
present model (full line) with the following values of the parameter 
$b=\beta_3^\text{w}/\beta_2^\text{w}$: 0.81
 for $^{224}$Ra, 0.68 for  $^{226}$Ra, and 0.85
for  $^{224}$Th.
The predictions of the X(5) model (dotted lines) and for a rigid 
reflection-asymmetric rotor (dashed-dotted)
are also shown for comparison.}
\label{F:5}
\end{figure*}

Now, as a first step, we can evaluate, as a function of $b$, the 
level energies in the ground--state band and deduce the best value 
of the parameter from a comparison
with experimental results (fig.~\ref{F:5}). 
To proceed, we must do some assumptions on the behavior of axial
and non axial modes of deformation. We will assume that
\begin{itemize}
\item Our choice of variables corresponds to independent degrees of
freedom.
\item Non-axial vibrations are confined to their lowest stationary 
state.
\item An approximation similar to that of the X(5) model is valid for
the differential equations of all non-axial amplitudes: {\sl i.e.},
the differential equation in $\beta_2,\ \beta_3$ can be approximately
decoupled from those concerning the other degrees of freedom.
\end{itemize}
Therefore, the complete wavefunction $\Psi$ of Eq.~\ref{E:5.0}
can be factorized in three parts, as in Eq. 30 of I:
\begin{equation}
\Psi=\Psi_0(\beta_2,\beta_3)\  \Psi_1\ Y_{JM}(\hat{\Omega})
\label{E:5.0}
\end{equation} 
where the function $\Psi_1$ depends on the deformation variables
 different from $\beta_2,\ \beta_3$.

From the Eq.~\ref{E:2.7} we know that also the determinant $G$
is factorized in the same way.
Then, the differential equation for $\beta_2 ,\ \beta_3$ takes the 
form\\[1mm]
\begin{eqnarray}
&&\hspace*{-4mm}
\left\{ G_0^{-1/2}\!\left[ \frac{\partial}{\partial 
\beta_2}\!\left( G_0^{1/2}
\!\frac{\partial}{\partial \beta_2 }\right)\!+\! 
\frac{\partial}{\partial \beta_3}\!\left( G_0^{1/2} 
\frac{\partial}{\partial \beta_3 }\right) \right] \right. \ \\
&&\hspace*{-4mm}\left. +\ \epsilon - V(\beta_2,
\beta_3) - \frac{J(J+1)}{3(\beta_2^2+2\beta_3^2)} 
\right\} \Psi (\beta_2, \beta_3) = 0 \nonumber
\label{E:5.1} 
\end{eqnarray}
This equation can be somewhat simplified with the substitution
\begin{eqnarray}
\Psi_0(\beta_2,\beta_3) = g^{-1/2}\ \Phi (\beta_2,\beta_3)
\label{E:5.2}
\end{eqnarray}
where $g\propto G_0^{1/2}$, to obtain
\begin{eqnarray}
\bigg\{
\frac{\partial^2}{\partial \beta_2^2} 
\!&\!+\!&\!\frac{\partial^2}{\partial \beta_3^2}
+ \epsilon - V(\beta_2,\beta_3)-\frac{J(J+1)}{3(\beta_2^2+2\beta_3^2)}
\nonumber \\
&\!+\!&\! {V_\text{g}(\beta_2,\beta_3)} \bigg\}\ \Phi (\beta_2,\beta_3) =0
\label{E:5.3} 
\end{eqnarray}
with
\begin{eqnarray}
V_\text{g}\!=\!\frac{1}{4g^2}  \bigg[ \left( \frac{\partial g
}{\partial \beta_2} \right) ^2\!+
\!\left( \frac{\partial g}{\partial \beta_3} \right) ^2 \bigg]
\!- \frac{1}{2g} \bigg[ \frac{\partial^2 g}{\partial \beta_2^2}
\!+\!\frac{\partial^2 g}{\partial \beta_3^2} \bigg]
\label{E:5.4} 
\end{eqnarray}

With the choice of $f(\beta_2, \beta_3)$ given in the Eq.~\ref{E:2.4},
from Eq.~\ref{E:2.4b} one obtains 
\begin{eqnarray}
g\propto \frac{(\beta_2^2\!+\beta_3^2)(\beta_2^2\!+\!2
\beta_3^2)^2}{(\beta_2^2+5\beta_3^2)} .
\label{E:5.5}
\end{eqnarray}
and, for  $|\beta_3| \ll \beta_2$,
$ g \propto \beta_2^4 \big[
1\!+\!4(\beta_3/\beta_2)^4\!+...\big] $.
Therefore, the first and second derivative of $g$ with respect to
$\beta_3$ tend to zero when $|\beta_3| \ll \beta_2$
and, at the limit  $\beta_3\rightarrow 0$, ${V_\text{g} = -2}$ as in the
original Bohr model.
\begin{figure}[b]
\includegraphics[height=7.5cm,angle=90,clip,bb=0 45 510 790]
{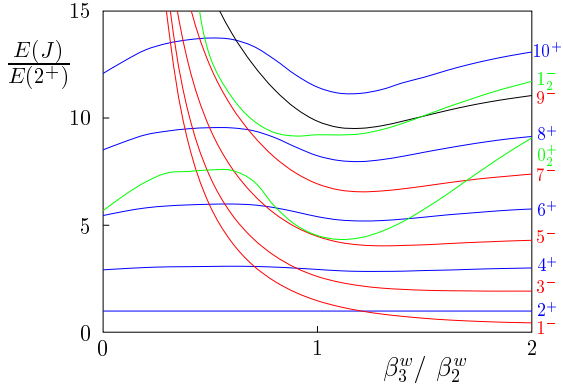}
\caption{\label{F:3.1} (Color online) Calculated energies of
excited levels of the ground and first excited band, in units of
$E(2^+_1)$, as a function of the ratio $b=\beta_3^w / \beta_2^w$.
}
\end{figure}
With the substitution $\Psi_0 =g^{-1/2}\Phi$, and assuming 
$V(\beta_2,\beta_3)=0$ inside the potential well and $=+\infty$
outside, the differential
equation to be solved takes the form
\begin{eqnarray}
\left[ \frac{\partial^2}{\partial
\beta_2^2}+\frac{\partial^2}{\partial \beta_3^2}+\epsilon 
+V_\text{g}(\beta_2,\beta_3) \right] \Phi (\beta_2,\beta_3) = 0
\label{E:5:6}
\end{eqnarray}
with $V_\text{g}$ given in the Eq.~\ref{E:5.4} 
and $\Phi=0$ on the contour of the potential well.
The numerical integration has been performed with the finite
difference method. 
Namely, the space is discretized on a rectangular
lattice and values of $\Phi$ at the lattice centers are taken as
independent variables. 
In the place of second derivatives, the ratios
of finite differences are  used: {\it e.g.},
\begin{eqnarray*}
\left( \frac{\partial^2 \Phi}{\partial \beta_2^2}
\right)_{x,y}
& \Rightarrow &
\frac{\Phi(x+\Delta_x,y)-2\ \Phi(x,y)
+\Phi(x-\Delta_x,y)}{ \Delta_x^2} 
\label{E:2.42}
\end{eqnarray*}
As 
$\Phi(\beta_2,\beta_3)=(-1)^J
\Phi(\beta_2,-\beta_3)$,
 it is enough to consider only
the region $\beta_3 >0$.
The lattice centers are chosen as
$\beta_2=k_2 \Delta_x,
\phantom{m}
\beta_3=(k_3-1/2) \Delta_y$
with $k_2=1...n_2$,  $k_3=1...n_3$, and 
$\Delta_x=\beta_2^w / (n_2+1)$, $\Delta_y=2\beta_3^w / (2n_3+1)$.
The integration region is the upper rectangle with 
$0<\beta_2<\beta_2^w$, $0<\beta_3<\beta_3^w$. At the upper and lateral
borders of the rectangle, the value of the eigenfunction must be zero.

The boundary conditions at $\beta_3=0$ are not specified, but to
evaluate the approximate derivatives with respect to $\beta_3$ it
is enough to consider the value of $\Phi$ at the line of centers 
immediately below zero, where they are either equal or opposite to the
corresponding ones at $\beta_3=\Delta_y/2$ according to the even or 
odd value of $J$.

The number of centers internal to the integration region -- and
therefore the number of independent values of $\Phi$
-- is now $N=n_2 \cdot n_3$,
and we obtain a finite dimensional $N\times N$ Hamiltonian matrix.
This Hamiltonian has been diagonalized with the Implicitly Restarted 
Arnoldi -- Lanczos method, using the ARPACK package~\cite{arpack}.

In the fig.~\ref{F:3.1}, calculated values of the excitation energies 
(in units of $E(2^+_1$)\ ) are depicted as a function of the ratio
$b=\beta_3^w/\beta_2^w$. At the limit for $\beta_3^w\rightarrow 0$,
the curves corresponding to even $J$ and $\pi$
tend to the X(5) values, as expected. With increasing $b$,
at the beginning these curves deviate substantially from the X(5) 
limit, but they come closer to the initial values for $b\approx 1$.
In this region it is possible to find a good fit of the ground--state
band of $^{224}$Ra and of $^{224}$Th, for $b=0.81$ and  for $b=0.85$,
respectively (Fig.~\ref{F:5}).

Moreover, a rather good fit of the ground--state band of $^{226}$Ra
is obtained with $b=0.68$, {\it i.e.} close to the maximum of the curves
for even parity and spin.

We can observe that, with our choice of the parameter $b$, the 
calculated $1^-$ level is always somewhat lower than the experimental 
one (Fig.~\ref{F:5}).
This fact can be related to the inclusion, in the potential well, of
a region where $\beta_3$ remains large while $\beta_2$ tends to zero.
Actually, the wavefunction of the first $1^-$ level extends appreciably
in this region, at variance with other levels of the ground-state band. 

It would be of great interest, of course, to extend the comparison to
the lowest excited band with $K=0$ 
(the $s=2$ band in the X(5) model notations). 
Unfortunately, in $^{224}$Th no excited $0^+$ level is known.
The non--yrast level schemes of $^{224,226}$Ra will be discussed in 
the following Section~\ref{S:4.2}
\begin{figure*}[!t]
\centering
\subfigure{
\includegraphics[height=7.5cm,angle=90,clip,bb=15 80 400 762]
{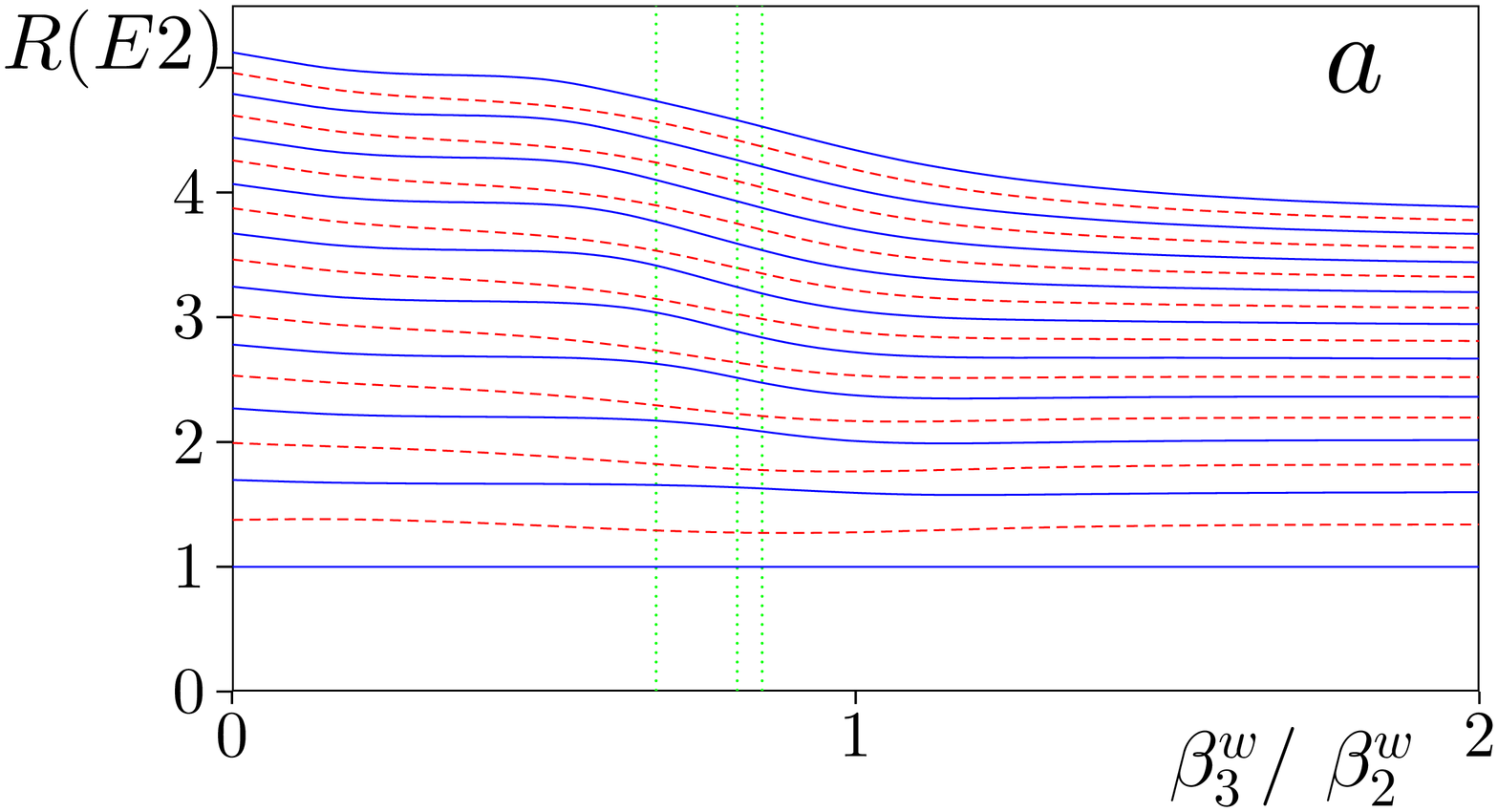}
}
\hspace{5mm}
\subfigure{
\includegraphics[height=7.5cm,angle=90,clip,bb=45 80 430 762]
{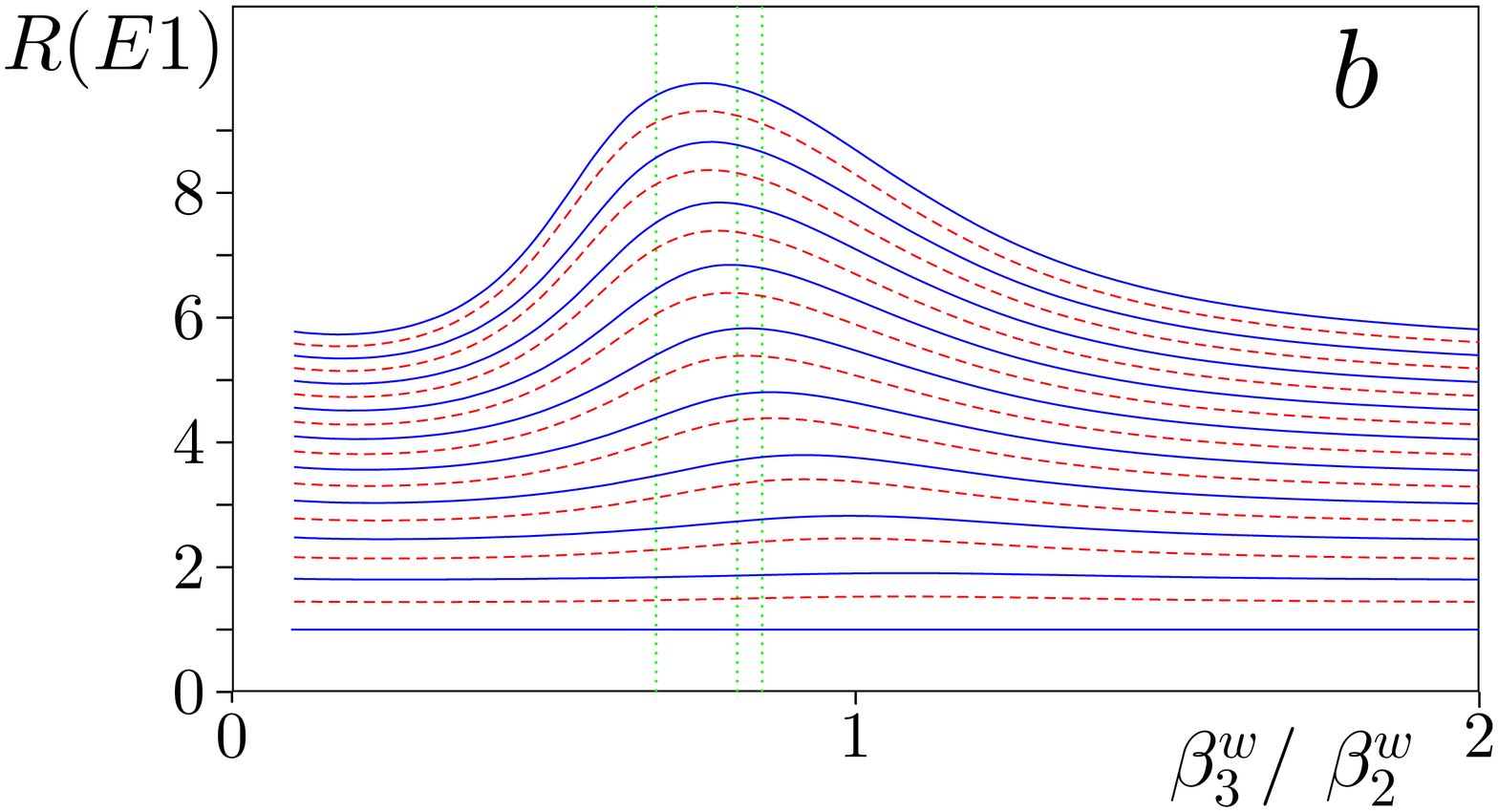}}
\caption{\label{F:7} (Color online)
Absolute values of the ratios of reduced matrix elements of the 
electromagnetic transition operators, $R($E$L,J_i)=
\mathcal{M}($E$L,J_i\rightarrow J_f))/\ \mathcal{M}($E$L,L\rightarrow 0)$,
with $J_f=J_i-L$, as a function of the parameter $b=\beta_3^w/ \beta_2^w$.
Part $(a)$: E2 transitions; solid lines: even $J_i^+ \rightarrow J_f^+$, 
starting with $2^+\rightarrow 0^+$ (from the bottom);
dashed lines: odd $J_i^- \rightarrow J_f^-$, 
starting with $3^-\rightarrow 1^-$.
Part $b$: E1 transitions; solid lines: odd $J_i^- \rightarrow J_f^+$, 
starting with $1^-\rightarrow 0^+$ (from the bottom);
dashed lines: even $J_i^+ \rightarrow J_f^-$, 
starting with $2^+\rightarrow 1^-$.
The vertical lines correspond to the adopted values of the parameter
for $^{226}$Ra, $^{224}$Ra and $^{224}$Th ($b=0.68,\ 0.81$ and $0.85$, 
respectively).
}
\end{figure*}

\begin{figure}[!b]
\centering
\includegraphics[height=11.cm,clip,bb=206 440 432 763]
{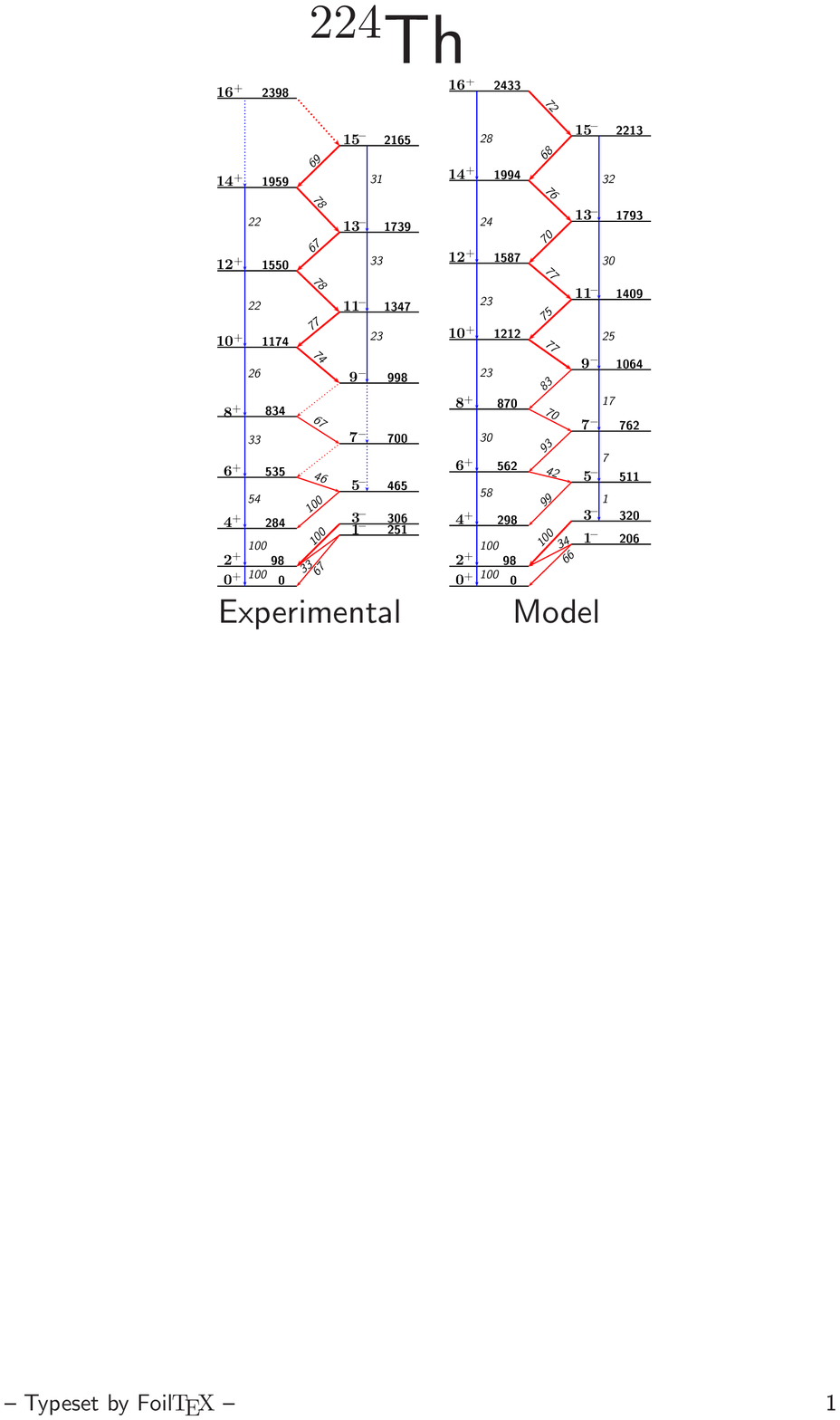}
\caption{\label{F:Th} (Color online) Level scheme of $^{224}$Th, compared 
with the model predictions for $b=0.85$. Calculated and
experimental branching ratios are reported for each level.
Experimental data are taken from
the NNDC tabulation~\cite{nndc}.
Experimental branching ratios from 
the $7^-$ and $9^-$ levels are not known. 
Theoretical values of the level energies (in keV) are normalized to that 
of the $2^+_1$ level; those of the branching ratios
are deduced from the matrix elements of Table~\ref{T:30}
with the experimental values of the transition energies.
Calculated branches lower than 1\% are not shown. 
}
\end{figure}

\subsection{Electromagnetic transition probabilities}
Another important test for the model is provided by the E2 transition
probabilities. The available experimental information on $B$(E2)
values is scarce (only two transitions in $^{224}$Ra and $^{226}$Ra,
one in $^{224}$Th), but we hope our work can stimulate interest for new
experimental investigations.
The reduced matrix element of 
the quadrupole transition operator ${\cal M}$(E2) 
between the states $|s,K=0,J \big>$ and $|s^\prime,K=0,J^\prime \big>$
can be evaluated as
\begin{equation}
\big( s J |\!|{\cal M}{\rm(E2)} |\!|s^\prime  J^\prime \big) = C_2
\big< s J | \beta_2 | s^\prime J^\prime \big> \ (J |\!| Y_2 |\!| 
J^\prime \big)\ ,
\label{E:14b}
\end{equation}
with $\big< s, J | \beta_2  |s^\prime ,  J^\prime \big> =
\int \Psi_{ s J} \beta_2 \Psi_{s^\prime ,
J^\prime}\ {\rm d}\tau$ and $C_2$ constant.
The volume element ${\rm d}\tau$, in our non-cartesian coordinates, is
the product of the differentials of the coordinate variables
{\em multiplied by} $g=G^{1/2}$, with $G$ the determinant of the matrix of
inertia $\cal G$.
In our assumptions, the integrals over all variables apart from
$\beta_2$ ad $\beta_3$ are independent from one another and from the
integral over ${\rm d}\beta_2{\rm d} \beta_3$, and their result is $1$
(if the corresponding wavefunctions are properly normalized). As the 
electric dipole and quadrupole operators do not contain derivatives,
we can exploit the substitution defined in Eq.~\ref{E:5.2} to express the 
remaining integral as
\begin{equation}
\int\!\Psi_{ s J} \beta_2 \Psi_{s^\prime 	
J^\prime}\ {\rm d}\tau
= \int_0^{\beta_2^w}\!{\rm d}\beta_2 	
\!\int_{-\beta_3^w}^{\beta_3^w}\!{\rm d}\beta_3
\Phi_{sJ}\beta_2\Phi_{s^\prime J^\prime}\phantom{.}
\end{equation}	
This integral has been evaluated numerically, for values of 
$J \leq 18$, with  $J^\prime = J-2$ (and also with $J^\prime = J-1$).
The reduced matrix element over the angular coordinates has the form
\begin{eqnarray}
&&\hspace*{-6mm}\big( J ||Y_L || J^\prime \big) = \\
&&\hspace*{-6mm}\phantom{m}(-1)^J (4\pi )^{-1/2} 
\sqrt{(2J\!+\!1) (2L\!+\!1) (2J^\prime\!+\!1)}
\left( \!\begin{array}{lcr} J&L&J^\prime \\ 0&0&0 \end{array}\!\right)
\nonumber
\end{eqnarray}
Finally, the reduced transition probabilities from $J$ to $J^\prime$
are obtained as \\
$B({\rm E2},s J\rightarrow s^\prime J^\prime)=(2J+1)^{-1}  
\big( s J ||{\cal M}{\rm(E2)} | |s^\prime  J^\prime \big) ^2$.\\
The absolute values  of the ratios of E2 reduced matrix elements, 
$R_J(E2)=(J|\!|\mathcal{M}(E2)|\!|J-2) /
(2^+|\!|\mathcal{M}(E2)|\!|0^+)$, for transitions
within the positive-- and the
negative--parity part of the ground--state band, are depicted, as a
function of $b=\beta_3^w/\beta_2^w$, in the fig.~\ref{F:7}$(a)$.
Their limit at $\beta_3^w/\beta_2^w\rightarrow 0$ corresponds, as 
expected, to the X(5) value.

In addition to the in--band E2 transition, we have to consider
the E1 transitions between levels of opposite parity.
How to treat E1 transitions in the frame of 
the geometrical model is a big problem, as all E1 transition moments 
should vanish for a homogeneous fluid of constant charge density.
In this sense, E1 transitions are outside the Bohr geometrical model.
It is usual to assume a constant {\em electric polarizability}
of the nuclear matter\cite{bohr57,bohr58}
to obtain the E1 operator in the form 
\begin{equation}
\mathcal{M}{\rm (E1)}= C_1 \beta_2 \beta_3 Y_1\ .
\label{E:15}
\end{equation} 
 This {\em ansatz} should be validated by proper 
microscopic calculations.

Actually, such a calculation has been performed by Tsvenkov {\it et
al.}~\cite{tsvenkov02} for a number of Radium, Thorium and Uranium 
isotopes, in the frame of the Skirme--Hartree--Foch model.
The electric dipole moment turns out to be almost independent of the
angular frequency in a given isotope, but can change drastically 
(even in the sign) along the isotopic chain. 
The small value of the electric dipole
moments in $^{224}$Ra is correctly predicted by these calculations.

Values of the ratios of the E1 matrix elements, 
$R_J(E1)=(J|\!|\mathcal{M}(E1)|\!|J-1) /
(1^-|\!|\mathcal{M}(E1)|\!|0^+)$, 
 obtained with the standard form (Eq.~\ref{E:15}) of the 
E1 operator, are shown in the fig.~\ref{F:7}$(b)$. They reach a
maximum for $\beta_3^w/\beta_2^w$ somewhat below 1, {\it i.e.} just in
a region including the values assumed for $^{224}$Ra and $^{224}$Th
(0.81 and 0.85, respectively).
The calculated values of $(J_i|\!|\mathcal{M}(EL)|\!|J_f)$
for E2 and E1 transitions in the ground state bands  of
$^{224}$Ra ($b=0.81$), $^{226}$Ra ($b=0.68$) and $^{224}$Th ($b=0.85$)
are given in the upper part of Table~\ref{T:30}.
Values for the corresponding
intra-band transitions are very similar in the three cases, 
while the 
difference can be larger for the weak inter-band  transitions,
as shown in the lower part of the Table.
{
\setlength{\tabcolsep}{0pt}
\begin{table}[b]
\caption{\label{T:30} Calculated values of the
reduced matrix elements of E1 and E2 transitions in $^{224,226}$Ra 
and $^{224}$Th, normalized to those of the lowest lying transition of 
the same multipolarity.
}
\begin{ruledtabular}
\begin{tabular}{ccccccccc}
\multicolumn{4}{c}{\large $ \big( J|\!| \mathcal{M}$(E1)$|\!|J^\prime 
\big)$ }&&
\multicolumn{4}{c}{\large $ \big( J|\!| \mathcal{M}$(E2)$|\!|J^\prime 
\big)$ }\cr
\cline{1-4}\cline{6-9}\\[-2.5mm]
Trans.&$^{224}$Ra&$^{226}$Ra&$^{224}$Th&&Trans.&$^{224}$Ra&$^{226}$Ra&%
$^{224}$Th\cr
\hline
$1^-\!\leftrightarrow\!0^+$&100&100&100&&$2^+\!\leftrightarrow\!0^+$&%
100&100&100\cr
$2^+\!\leftrightarrow\!1^-$&149&147&150&&$3^-\!\leftrightarrow\!1^-$&%
127&129&127\cr
$3^-\!\leftrightarrow\!2^+$&187&184&187&&$4^+\!\leftrightarrow\!2^+$&%
164&166&163\cr
$4^+\!\leftrightarrow\!3^-$&238&228&241&&$5^-\!\leftrightarrow\!3^-$&%
179&182&178\cr
$5^-\!\leftrightarrow\!4^+$&273&262&276&&$6^+\!\leftrightarrow\!4^+$&%
211&217&209\cr
$6^+\!\leftrightarrow\!5^-$&333&311&338&&$7^-\!\leftrightarrow\!5^-$&%
223&230&221\cr
$7^-\!\leftrightarrow\!6^+$&371&346&376&&$8^+\!\leftrightarrow\!6^+$&%
252&264&248\cr
$8^+\!\leftrightarrow\!7^-$&435&402&439&&$9^-\!\leftrightarrow\!7^-$&%
264&274&261\cr
$9^-\!\leftrightarrow\!8^+$&476&438&480&&$10^+\!\leftrightarrow\!8^+%
\phantom{1}$&289&304&284\cr
$10^+\!\leftrightarrow\!9^-\phantom{1}$&538&500&539&&$%
11^-\!\leftrightarrow\!9^-\phantom{1}$&303&315&299\cr
$11^-\!\leftrightarrow\!10^+$&582&538&582&&$12^+\!\leftrightarrow\!10^+%
$&325&342&320\cr
$12^+\!\leftrightarrow\!11^-$&640&604&635&&$13^-\!\leftrightarrow\!11^-%
$&340&354&336\cr
$13^-\!\leftrightarrow\!12^+$&685&644&680&&$14^+\!\leftrightarrow\!12^+%
$&360&377&354\cr
$14^+\!\leftrightarrow\!13^-$&738&709&730&&$15^-\!\leftrightarrow\!13^-%
$&376&390&371\cr
$15^-\!\leftrightarrow\!14^+$&783&750&775&&$16^+\!\leftrightarrow\!14^+%
$&393&411&388\cr
$16^+\!\leftrightarrow\!15^-$&833&812&822&&$17^-\!\leftrightarrow\!15^-%
$&410&424&405\cr
$17^-\!\leftrightarrow\!16^+$&878&854&867&&$18^+\!\leftrightarrow\!16^+%
$&426&443&421\cr
$18^+\!\leftrightarrow\!17^-$&925&912&912&&$19^-\!\leftrightarrow\!17^-%
$&442&457&437\cr
\hline
$0_2^+\!\leftrightarrow\!1_1^-$&\ 84&\ 42&150&&$%
0_2^-\!\leftrightarrow\!2_1^+$&$\phantom{1.}$8&\ 24&$\phantom{1.}$6\cr
$1_2^-\!\leftrightarrow\!0_1^+$&\ 31&\ 31&\ 31&&$%
1_2^-\!\leftrightarrow\!1_1^-$&\ 38&\ 37&\ 38\cr
$1_2^-\!\leftrightarrow\!2_1^+$&\ 49&\ 50&\ 47&&$%
1_2^-\!\leftrightarrow\!3_1^-$&\ 42&\ 43&\ 43\cr
$1_2^-\!\leftrightarrow\!0_2^+$&\ 22&\ 63&\ 21&&$%
2_2^+\!\leftrightarrow\!0_1^+$&\ 15&\ 17&\ 16\cr
$2_2^+\!\leftrightarrow\!1_1^-$&113&\ 86&111&&$%
2_2^+\!\leftrightarrow\!2_1^+$&\ 18&\ 25&\ 17\cr
$2_2^+\!\leftrightarrow\!3_1^-$&143&103&142&&$%
2_2^+\!\leftrightarrow\!4_1^+$&$\phantom{1.}$8&\ 32&$\phantom{1.}$4\cr
$2_2^+\!\leftrightarrow\!1_2^-$&\ 33&\ 71&\ 34&&%
$2_2^+\!\leftrightarrow\!0_2^+$&\ 91&\ 82&\ 92\cr
\end{tabular}
\end{ruledtabular}
\end{table}
}
\subsection{Comparison with experimental transition
probabilities}
{
\setlength{\extrarowheight}{4pt}
\begin{table*}
\caption{\label{T:e1} Experimental and calculated values of the ratios of 
reduced amplitudes of two E1 or two E2 transitions.
}
\begin{ruledtabular}
\begin{tabular}{lcccrrrrrrrrrrr}
&&&&\multicolumn{11}{c}{\large $(\ J_{\rm A}\ |\!|\mathcal{M}$(EL)$|\!|\ J_{\rm A}^\prime \ )$\ /
\ $(\ J_{\rm B}\ |\!|\mathcal{M}$(EL)$|\!|\ J_{\rm B}^\prime \ )$}\\[4pt]
\cline{6-15}\\[-4mm]
&\multicolumn{3}{c}{Transitions}&&
\multicolumn{2}{c}{\large $^{224}$Ra}&&\multicolumn{2}{c}{\large $^{226}$Ra}&&
\multicolumn{2}{c}{\large $^{224}$Th}\cr
\cline{2-4}\cline{6-7}\cline{9-10}\cline{12-13}
\phantom{i}&$J_{\rm A}\rightarrow J_{\rm A}^\prime$ &
&$ J_{\rm B} \rightarrow J_{\rm B}^\prime$&&Experim.&Crit.&&Experim.&Crit.&&Experim.&Crit.&&Rot.\cr 
\hline
E1&$1_1^-\!\rightarrow\!2_1^+$&&$1_1^-\!\rightarrow\!0_1^+$&&
$1.52\pm 0.14$ & 1.50&&$1.36\pm0.12$&1.47&&$1.49\pm0.26$&1.50&&1.42\cr
E1&$3_1^-\!\rightarrow\!4_1^+$&&$3_1^-\!\rightarrow\!2_1^+$&&
&&&$1.11\pm0.18$&1.24&&&&&1.15\cr
E2&$4_1^+\!\rightarrow\!2_1^+$&&$2_1^+\!\rightarrow\!0_1^+$&&
$1.60\pm 0.05$ & 1.63&&$\approx 1.76\phantom{22}$&1.66&&&&&1.60\cr
\hline
E2&$1_2^-\!\rightarrow\!3_1^-$&&$1_2^-\!\rightarrow\!1_1^-$&&
$0.71\pm 0.10$& 1.10&&&&&&&&--\ \cr
E1&$1_2^-\!\rightarrow\!2_1^+$&&$1_2^-\!\rightarrow\!0_1^+$&&
$1.49\pm 0.16$ & 1.57&&$1.24\pm0.09$&1.62&&&&&--\ \cr
E1&$2_2^+\!\rightarrow\!3_1^-$&&$2_2^+\!\rightarrow\!1_1^-$&&
&&&$1.29\pm0.08$&1.20&&&&&--\ \cr
\end{tabular}
\end{ruledtabular}
\end{table*}
}
{
\setlength{\extrarowheight}{2pt}
\begin{table*}
\caption{\label{T:3} Experimental and calculated values of the ratios of 
reduced amplitudes of E1 and E2 transitions from the same level (in units 
of their Weisskopf estimates). The columns of calculated values are
normalized to obtain the best fit to the experimental
values for the transitions within the ground--state band.
}
\begin{ruledtabular}
\begin{tabular}{lcccrrrrrrrr}
&&&\multicolumn{9}{c}{\large $\big[ \big( \mathcal{M}$(E1)/$\mathcal{M}_W$(E1)$ \big) 
/\big( \mathcal{M}$(E2)/$\mathcal{M}_W$(E2)$\big) \big] \times 10^3$}\\[4pt]
\cline{4-12}\\[-3mm]
&\multicolumn{2}{c}{\large Transitions}&\multicolumn{3}{c}{\large $^{224}$Ra}&\multicolumn{3}{c}{\large $^{226}$Ra}&
\multicolumn{3}{c}{\large $^{224}$Th}\cr
\cline{1-3}\cline{4-6}\cline{7-9}\cline{10-12}
&E1&E2&Experim.&Crit.&Rot.&Experim.&Crit.&Rot.&Experim.&Crit.&Rot.\cr 
\hline&$3_1^-\!\rightarrow\!2_1^+$& $3_1^-\!\rightarrow\!1_1^-$&
$0.69\pm 0.14\phantom{i}$& 0.55&0.57&\cr
&$5_1^-\!\rightarrow\!4_1^+$&$5_1^-\!\rightarrow\!3_1^-$&
$0.98\pm 0.26$\footnote{from NNDC~\cite{nndc} only. The $5^-\rightarrow 3^-$ (142 keV)
 $\gamma$ ray observed in the reaction data~\cite{cocks99} appears to be
contaminated by a close-lying transition from a different reaction, as it 
results from the intensity mismatch in the 
$5^-\rightarrow 3^-\rightarrow 2^+ (\rightarrow 1^-)$ cascade.} 
& 0.63 &0.60&$1.36\pm 0.23$\footnote{from Ref~\cite{ackermann93}. Not included in the fit.}
&2.13 &2.42\cr
&$6_1^+\!\rightarrow\!5_1^-$&$6_1^+\!\rightarrow\!4_1^+$&
&&&&&&$7.98\pm 1.17$&7.18&8.21\cr
&$7_1^-\!\rightarrow\!6_1^+$& $7_1^-\!\rightarrow\!5_1^-$&
$0.56\pm 0.09$
& 0.66&0.65 &$2.51\pm 0.15$&2.23 &2.56\cr
&$8_1^+\!\rightarrow\!7_1^-$ &$8_1^+\!\rightarrow\!6_1^+$&
$<1.22\phantom{9i}$&0.68&0.66 &&&
&$7.19\pm 0.72$& 7.86&8.50\cr
&$9_1^-\!\rightarrow\!8_1^+$&$9_1^-\!\rightarrow\!7_1^-$&
$<1.71\phantom{9i} $&0.71  &0.67 &$2.82\pm 0.34$&2.38&2.67\cr
&$10_1^+\!\rightarrow\!9_1^-\phantom{0}$&$10_1^+\!\rightarrow\!8_1^+\phantom{0}$
&&&&&&&$7.78\pm 0.43$& 8.41&8.67\cr
&$11_1^-\!\rightarrow\!10_1^+$ &$11_1^-\!\rightarrow\!9_1^-\phantom{0}$&&&&$2.76\pm 0.27$
&2.53&2.67 &$9.35\pm0.62$& 8.64&8.73\cr
&$12_1^+\!\rightarrow\!11_1^-$ &$12_1^+\!\rightarrow\!10_1^+$&&&&$2.85\pm 0.25$
&2.62&2.68 &$9.06\pm0.47$&8.83&8.78\cr
&$13_1^-\!\rightarrow\!12_1^+$ &$13_1^-\!\rightarrow\!11_1^-$&&&&$2.15\pm 0.29$
&2.70&2.70 &$8.45\pm0.42$ & 9.01&8.82\cr
&$14_1^+\!\rightarrow\!13_1^-$ &$14_1^+\!\rightarrow\!12_1^+$&&&&$2.58\pm 0.17$
&2.79&2.71 &$9.84\pm 0.51$&9.15&8.86\cr
&$15_1^-\!\rightarrow\!14_1^+$ &$15_1^-\!\rightarrow\!13_1^-$&&&&$2.53\pm 0.17$
&2.94&2.73 &$9.69\pm0.65$ & 9.29&8.89\cr
&$17_1^-\!\rightarrow\!16_1^+$ &$17_1^-\!\rightarrow\!15_1^-$&&&&$2.78\pm 0.43$
&2.98&2.74 &$10.47\pm 1.34$&9.52&8.92\cr
&$18_1^+\!\rightarrow\!17_1^-$&$18_1^+\!\rightarrow\!18_1^+$ &&&&$3.22\pm 0.21$
&3.05& 2.74\cr
\hline
&\multicolumn{4}{l}{$\chi^2\ / \ n\ (with\ n\ =\ 8)$}&&&2.13&1.34&&1.17&2.03\cr
&\multicolumn{4}{l}{$Confidence\ level\ (\%)$}&&&$<$5 &18&&31&$<$5\cr
\end{tabular}
\end{ruledtabular}
\end{table*}
}
The Table~\ref{T:e1} shows a few values of the ratio of reduced 
matrix elements for transitions of the same multipolarity, that can 
be deduced from the available experimental information. 
In the same Table, the corresponding values calculated with the 
present model are also shown (columns ``Crit.''), together
with the ones expected for a reflection-asymmetric rigid rotor
(``Rot.'').
 
The most direct check of the model predictions would come from the
ratios of B(E2) values in the ground--state band. 
This is possible only in $^{224}$Ra, and
only for the decays of the lowest $2^+$ and $4^+$ levels.
With the experimental values reported in the NNDC
tabulation~\cite{nndc}, $B($E2$,2^+\rightarrow 0^+)=97\pm 3$~W.u. 
and $B($E2$,4^+\rightarrow 2^+)=138\pm 8$~W.u.,
the experimental value of the ratio is $1.42\pm 0.09$, to be compared 
with the value 1.41 that is obtained from the calculated matrix 
elements of Tables~\ref{T:30} and~\ref{T:e1} (for $b=0.81$).
We remind that, in the X(5) model~\cite{iachello01}, this ratio would be 1.59. 
For  $^{226}$Ra, the lifetime of the $4^+$ state is known, but for 
the first excited state only an approximate value (without error 
estimate) is reported. Also in this case, the deduced ratio is 
consistent with the theoretical estimate (see Table~\ref{T:e1}).
These results is encouraging,
but would obviously need to be validate by a more extensive check,
involving higher--lying levels, which, at the moment, is not
possible. 

A comparison of the two E1 transition from the lowest level $1^-$  
to the
$0^+$ and to the $2^+$ states is possible for the three isotopes, as
well as for the E1 branches from the $3^-$ in $^{226}$Ra.
All these amplitude ratios for transitions within the ground--state band, 
shown in the upper part of Table~\ref{T:e1}, are in very good agreement 
with the calculated values. 
However, they are not significantly different from those expected for a 
rigid asymmetric rotor (as shown in the last column of Table~\ref{T:e1})
and also from those reported by Lenis and Bonatsos~\cite{lenis06} on the
basis of a rather different model.
\begin{figure}[!b]
\centering
\includegraphics[height=7.cm,clip,bb=20 320 592 815] 
{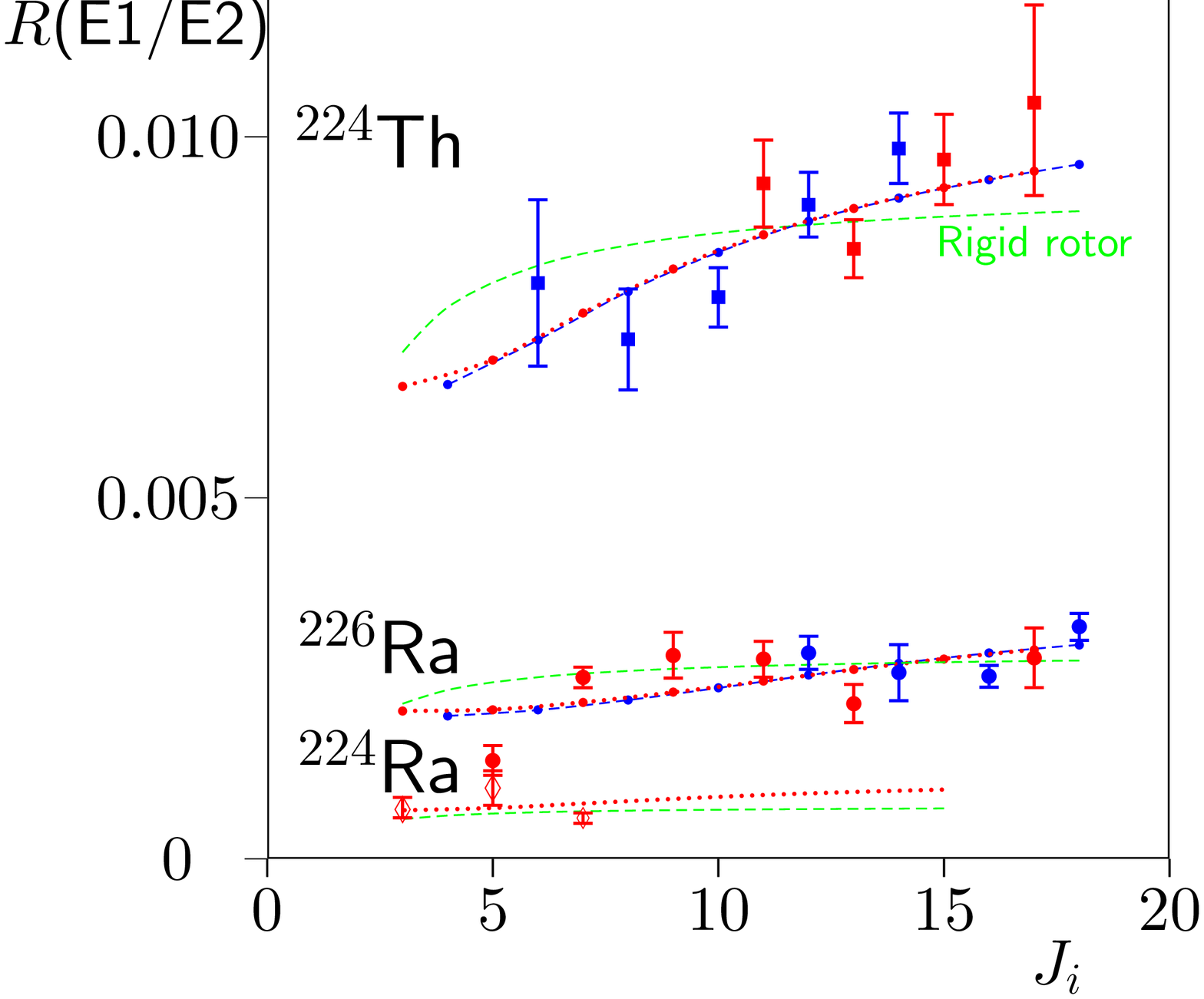}
\caption{\label{F:e1e2} (Color online) Ratios of the absolute value of
the transition matrix elements (normalized to the Weisskopf Unit)
for E1 and E2 transitions in the g.s bands of $^{224,226}$Ra and
$^{224}$Th, from $J_i$ to $J_i\!-\!1$ and $J_i\!-\!2$, respectively:
$R_J(E1/E2)=$
$(J_i|\!|[\mathcal{M}(E1)|\!|J_i\!-\!1)/\mathcal{M}_W(E1)]
/[(J_i|\!|\mathcal{M}(E2)|\!|J_i\!-\!2)/\mathcal{M}_W(E2)]$.
The dotted lines join the calculated values of the ratio (normalized to 
obtain the best fit with the ensemble of experimental values). 
The dashed lines join the values expected for a rigid rotor.
The corresponding values deduced from the parameter free 
model of Ref.~\cite{lenis06} are (apart for a possible
staggering between even and odd $J_i$) almost identical to the
rotational ones for large values of $J_i$ ($J_i>7$) and, for decreasing
values of $J_i$, their trend reaches a minimum around $J_i=6$ and 
then increases slightly at lower values of $J_i$.
}
\end{figure}
We note that, when the transitions to be compared have the
same multipolarity, the model predictions are parameter free, or --
more exactly -- only involve the model parameter $\beta_3^w/\beta_2^w$.

Instead, when the comparison concerns
the ratios of the reduced matrix elements for E1 and E2 transitions 
deexciting the same level, the model predictions 
include a further normalization factor (the ratio of constants $C_1$ 
and $C_2$ of Eq.s~\ref{E:14b},\ref{E:15})
which needs to be determined from the experimental data.
This comparison is therefore less direct, but it is perhaps 
more significant, as we shall see in the following.

Results concerning the E1/E2 branches in the ground--state band are shown
in the Table~\ref{T:3} and also depicted in Fig.~\ref{F:e1e2}.
Experimental values of E1/E2 branching ratios in $^{224,226}$Ra and 
$^{224}$Th include those given in the NNDC 
tabulation~\cite{nndc,ackermann93} and later results 
from Ref.~\cite{cocks99}. From these branching ratios we have deduced 
the absolute ratios -- given in the ``Exp.'' columns of Table~\ref{T:3} 
-- of the reduced matrix elements of  E1 
and E2 transitions, each of which is expressed in units of the 
corresponding Weisskopf estimate, 
$\mathcal{M}_W(EL)= 
(4\pi)^{-1/2}[3/(L+3)]\ (1.2 A^{1/3})^L$~e~fm$^L$.
In the same table are also shown the results of the model
calculation at the critical point (Crit.), which have been normalized to 
obtain the best fit with  the experimental values within the 
ground--state band of each nucleus.
Values expected for a rigid asymmetric rotor (Rot.), normalized in the 
same way, are also shown.
The $^{226}$Ra point at $J_i=5$ which, according to the authors 
themselves~\cite{ackermann93},
could be considered as a lower limit, has not been included in the fits.

For the ground--state band of $^{224}$Th (Fig.~\ref{F:Th},\ref{F:e1e2}), 
we find a satisfactory agreement between
the experimental values and the model predictions. In this case we have
enough data to perform a $\chi^2$ test of goodness of fit, and we obtain
$\chi^2/N= 1.17$ with $N=8$ degrees of freedom, corresponding to 
a confidence level of 31\%. A fit with the rigid--rotor values would give 
a much larger value $\chi^2/N=2.03$, and a confidence level below 5\%.
\begin{figure*}
\centering
\includegraphics[height=\linewidth,clip,angle=90,bb=98 7 475 800]
{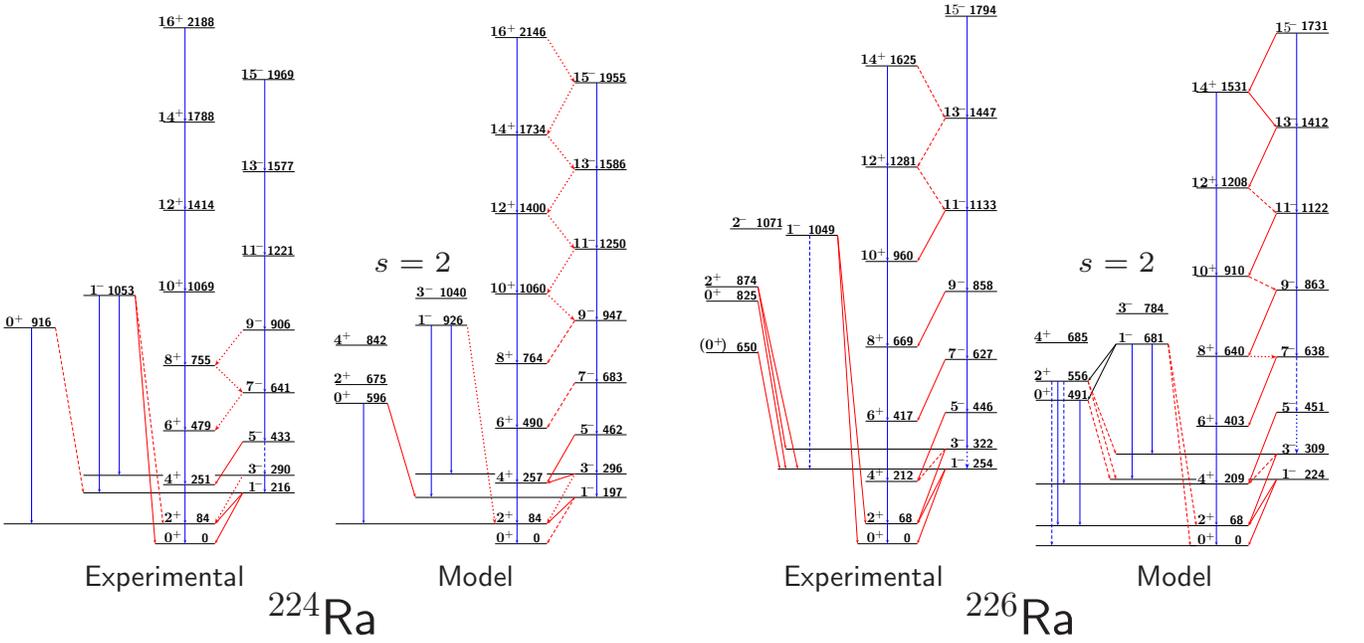}
\caption{\label{F:6} (Color online) 
Partial level schemes of $^{224}$Ra and $^{226}$Ra, with the 
experimentally observed $\gamma$ transitions, compared with the 
results of model calculations (with
$b=0.81$ and $=0.68$, respectively). Theoretical level energies (in keV)
are normalized to that of the first excited level.
Experimental energies for the lower levels of $^{224}$Ra  are taken 
from the NNDC tabulation~\cite{nndc}, those of the $10^+$, $12^+$ 
and higher levels are deduced from the $\gamma$--ray energies given 
by Cocks {\it et al.}~\cite{cocks99}. For $^{226}$Ra, those of levels 
up to $5^-$ are taken from NNDC or ref.~\cite{ackermann93},
those of higher levels from Cocks {\it et al.}\cite{cocks99}.
Gamma branches lower than 5\% 
(or reported as upper limits) are shown 
as dotted lines, those between 5\% and 25\% as dashed lines. Calculated 
branches lower than 1\% are not shown. 
For a comparison of experimental  E1/E2 branches with the model prediction
at the critical point, see Table~\ref{T:3} and Fig.~\ref{F:e1e2}.
}
\end{figure*}
Also for the ground--state band of $^{224}$Ra, the few available
experimental values (or limits) are not far from the results
of the model, but more experimental data would be necessary for a 
significant comparison. 
Actually, as it was soon recognized~\cite{marten90,butler91,egido91} 
the E1 transitions in $^{224}$Ra 
are rather weak compared to other nuclei in this region, and in 
particular their strengths are two orders of magnitude smaller than 
the corresponding ones in $^{224}$Th.
 
Instead, experimental values for $^{226}$Ra deviate significantly from 
the model predictions and approach those expected for a rigid rotor. 
This fact, combined with the slight upward deviation of level energies 
from the calculated curve for $J>14$, suggests that the critical
point of the phase transition in the Ra isotopic chain can be situated 
somewhere below $A=226$, and probably close to $A=224$.

\subsection{\label{S:4.2}The first excited $K=0$ band}

As anticipated in Section ~\ref{S:4.1}, no experimental information is 
available for non yrast levels of $^{224}$Th. 
For $^{224,226}$Ra isotopes, a few non-yrast
levels are known from $\beta^-$ decay of $^{224,226}$Fr, from $\alpha$ 
decay of $^{228,232}$Th or from the $^{226}$Ra(t,p) reaction.
Unique assignments of the spin and parity have been reported only for 
part of them. 
Some of these levels, which could be considered as members of the
excited  $K=0$ band (the $s=2$ band, in~the X(5) expression) 
 are reported, together with those of the yrast 
band, in Fig.~\ref{F:6}, where also the main decay branches are 
indicated.
In the same figure, the model predicted levels, and their 
expected $\gamma$ branches, are also shown. 

We can immediately observe that non yrast levels predicted by the model
are always lower than the experimental ones (but a comparably large
discrepancy is observed also in the $s=2$ band of X(5) 
nuclei~\cite{casten01,kruecken02,tonev04}).
In the lower part of Table~\ref{T:e1},
the calculated amplitude ratios for transitions from the excited $K=0$ 
band are compared with the corresponding experimental ones, if the levels 
$0^+_2$ and $1^-_2$  shown in the Fig.~\ref{F:6} are interpreted as 
belonging to it.
Only the ratio of the two E1 transitions from the $1^-_2$ level of
$^{224}$Ra and from the $2^+_2$ level of $^{226}$Ra are well consistent
with the calculated value, while the corresponding ratios for the two
E1 transitions from the $1^-_2$  level of $^{224}$Ra and
for the two E2 transitions from the $1^-_2$ level of $^{224}$Ra
seem to be significantly different 
from the model predictions  (although the latter is subject to 
a large uncertainty, due to the presence of a competing M1 component
in the $1^-_2\rightarrow 1_1^-$ transition).

As for the E1/E2 ratios for inter-band transitions, it is not obvious
that the value of the parameter $C_1/C_2$ ought to be the same as for
transitions within the ground--state band, but if we assume to be so,
the E1/E2 ratios in the decay of the $1^-_2$ level of $^{226}$Ra 
differ by a factor
of 2 from the calculated values: the ratios to the E2 amplitude
$1^-_2\rightarrow 3^-_1$, with the normalization used in the the 
Table~\ref{T:3},
are $(0.75\pm0.7)10^{-3}$ for the $1^-_2\rightarrow 0^+_1$ E1 transition
and $(1.05\pm 0.07)10^{-3}$ for the $1^-_2\rightarrow 2^+_1$, to be 
compared with the theoretical values $0.29\cdot  10^{-3}$ and 
$0.46\cdot 10^{-3}$, respectively. 
  
Therefore, if the first two levels of the excited $s=2$ band 
are tentatively identified with the $0^+_2$
and $1^-_2$ levels of $^{224}$Ra, their properties 
are not so well accounted for. For this fact, one can 
hypothesize different explanations. 
First, we remark that the identification of these levels 
as members of the $\beta$ band can be put in discussion.
Actually, the $0_2^+$ level could result from other
(collective or non collective) modes of excitation,
as, {\it e.g.}, pairing vibration~\cite{friedman74,vanrij72},
while the $1^-_2$ could correspond to (or be mixed with) the band head of 
the $K^\pi=1^-$ band. Otherwise, the observed
disagreement could indicate that our model is unable to
correctly predict states outside the ground--state band,
in particular if they are not far from
levels of the non axial modes having the same $J^\pi$.
The simultaneous investigation of axial and non axial modes,
as it has been performed, via the Extended Coherent State
model, in the ref.s~\cite{raduta03,raduta06b,raduta06c,radutapredeal},
is outside our present possibilities.
\section{Conclusions}
An extension of Iachello's X(5) model to 
the axial quadrupole + octupole deformation has been developed
with the formalism introduced in our previous paper I~\cite{bizzeti04}.
Assuming that both $\beta_2$ and $\beta_3$ can
vary within a two--dimensional well with rectangular 
borders, and with a proper determination of a free
function of the model, the results are found to converge to those
of X(5) when the interval available for $\beta_3$ tends to zero.
The formalism is therefore suitable to describe the
critical point of phase transitions involving at the same
time the axial quadrupole and octupole deformation.

As anticipated in I, the principal aim of this
second part of our work 
was the description of the transitional nuclei 
$^{224}$Ra and $^{224}$Th, which were proposed 
to be close to such a critical point.

Actually, in spite of the admittedly crude schematization of the
bidimensional potential, the relative values of the
excitation energies of levels 
(of positive and negative parity) in the ground--state 
bands of both $^{224}$Ra and $^{224}$Th
are satisfactorily reproduced by adjusting the only
available parameter (the aspect ratio $b=\beta_3^w/\beta_2^w$
of the potential well). A good agreement is obtained with $b=0.81$
for $^{224}$Ra and with $b=0.85$ for $^{224}$Th.
Moreover, a good agreement is also obtained for the first part of
the ground--state band of $^{226}$Ra with a lower value of the parameter,
$b=0.68$: only above $J=14$ the experimental points deviate slightly from
the calculated values, in the direction of the rigid--rotor curve
(Fig.~\ref{F:5}).

The (few) known ratios of transition strengths in the ground state band,
for electromagnetic 
transition of equal multipolarity (either E2 or E1) 
are in agreement with the model predictions.
Unfortunately, only in a few cases the ratio of the 
reduced strengths for transitions of equal multipolarity
can be deduced from the experimental data (see Table~\ref{T:e1})
and in these cases the values expected at the critical point
are not very different from those of the rotational model. 

In some more cases, the relative strength of two
transitions of different multipolarity (E1 and E2),
coming from the same level, can be deduced from the measured
branching ratio. The comparison with the model requires in
this case one more parameter, which has been determined
by a best--fit procedure (see Table~\ref{T:3} and Fig.~\ref{F:e1e2}).
But, in this case, the expected trend at the critical--point is 
significantly different from that of a rigid rotator.

The calculated critical--point values of the ratios E1/E2
are in a rather good agreement with the experimental results 
in the case of $^{224}$Th (Fig.~\ref{F:e1e2}),
while for $^{226}$Ra the
trend of empirical values is closer to the one expected
for a rigid rotor. For $^{224}$Ra, the E1
transitions are very weak and experimental data are too scarce to permit 
a significant comparison with the model predictions.

New and more
extensive measurements of the transition strengths 
either in $^{224}$Ra or $^{224}$Th would be highly
desirable, for a more significant test of the model.
\bibliography{biz_bib}
\end{document}